\documentclass[11pt,twoside]{article}

\usepackage{asp2006arxiv}
\usepackage{epsf}

\begin{document}
\title{LENS MAPPING OF DARK MATTER SUBSTRUCTURE WITH VSOP-2}
\author{Shigenori Ohashi,\altaffilmark{1} Masashi Chiba,\altaffilmark{1}
  and Kaiki Taro Inoue \altaffilmark{2}} 

\altaffiltext{1}{Astronomical Institute, Tohoku University, Aoba-ku, Sendai, 980-8578, Japan}
\altaffiltext{2}{Department of Science and Engineering, Kinki University, Higashi-Osaka, 577-8502, Japan}

\begin{abstract} 
Hierarchical clustering models of cold dark matter (CDM) predict that about 5\% - 10\% of a galaxy-sized halo with mass $ \sim 10^{12}$ solar masses ($M_{\odot }$) resides in substructures (CDM subhalos) with masses $\la 10^{8} M_{\odot }$. To directly identify such substructures, we propose to observe radio continuum emission from multiply imaged QSOs using VSOP-2 with a high angular resolution.
\end{abstract}

\section{Introduction}
The currently standard framework for understanding cosmological structure is based on hierarchical clustering of CDM in the accelerating Universe. Low mass systems collapse and form earlier and later they merge to form more massive gravitationally bound systems. The standard model is successful for explaining a wide variety of astrophysical observations such as the temperature fluctuations in the Cosmic Microwave Background and large-scale distribution of galaxies. Despite the great success of the concordance model, there are several open issues. One of them is so-called "the missing satellite problem". Some authors have argued that this discrepancy could be resolved by considering some suppressing process for star formation, such as gas heating by an intergalactic UV background. In whatever models relying on the suppression of galaxy formation, a typical galaxy-sized halo should contain numerous dark subhalos. It is thus important to constrain the presence or absence of many invisible satellites around a galaxy like the Milky Way in the form of dark subhalos.

To obtain the direct evidence for such dark subhalos, we propose to observe the multiply imaged QSOs with a high angular resolution. Irrespective of the presence or absence of associated luminous components, it is possible to probe these subhalos in a foreground lensing galaxy by means of gravitational lensing.

\section{Theory of Gravitational Lensing}
The phenomena that light rays are deflected by gravity is one of the consequences of Einstein's General Theory of Relativity. This is referred to as Gravitational lensing. Gravitational lensing is a powerful tool to determine the mass of lens objects which deflect and distort a source image in the background.

In a gravitational lens system, Einstein radius is a useful indicator of the scale of image distortion. In the case of an Singular Isothermal Sphere (SIS) with density distribution of $\rho \propto r^{-2}$, Einstein radius is written as,
\begin{equation}
 \theta_E \sim 1.5 \left(\frac{M_t}{5\times10^7 M_\odot}\right) \left({\frac{r_t}{1.0\mathrm{kpc}}}\right)^{-1}\left(\frac{D_{ls}/D_{os}}{0.5}\right) [\mathrm{mas}]
\end{equation}
where $r_t$ is a supposed tidal radius of an SIS lens and $M_t$ is a total mass enclosed within $r_t$. $D_{ls}$ and $D_{os}$ are angular diameter distances between lens and source and between observer and source, respectively.

Several observations of lens systems in a galactic scale have revealed that the density distribution of a galaxy lens, being composed of both baryonic and dark matter, is well represented by an SIS profile.
However, actual density profiles in subhalos are yet to be settled. According to N-body simulations based on the CDM model, i.e., without taking into account dissipative baryonic matter, all dark halos appear to show a shallower density profile at their central parts, expressed as $\rho \propto r^\gamma$ with $-1.5 < \gamma < -1$ (e.g., Navarro et al. 2004).
If this is the case, then a dark halo with this profile has a smaller Einstein radius than an SIS case, so the resulting lensing signal to be observed by VSOP-2 would be biased further in favor of more massive subhalos. How baryonic components, if there are any, affect a dark halo profile is yet uncertain: an adiabatic contraction of a halo as driven by gas cooling yields a steeper density profile, whereas starburst activity and associated heating by supernovae explosion may make a gravitational potential shallower owing to galactic wind.
Whatever effects of baryonic components are at work, an SIS profile would provide a possibly upper limit to the lensing signal of subhalos.

As is shown in equation (1), the size of an Einstein radius depends on the distance ratio, $D_{ls}/D_{os}$. We plot, in Figure 1, this dependence for a subhalo of $M_t = 5 \times 10^7 M_\odot$ and $r_t = 1.0$ kpc, using known lensing systems taken from CASTLES.
\begin{figure}[!ht]
\plotone{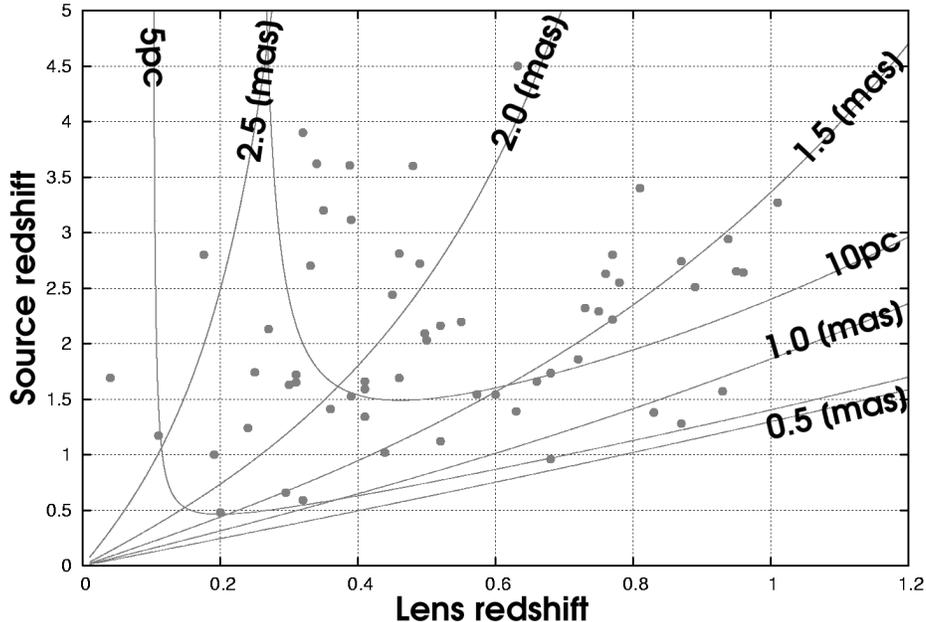}
\caption{The dependence of an Einstein radius on lens and source redshift in the case of $M_t=5\times 10^7 M_\odot$ and $r_t=1.0$kpc. Filled circles: known lensing systems taken from CASTLES. Solid lines: the size of an Einstein angle in units of mas or that of an Einstein raidus projected in a lens plane in units of pc.}
\end{figure}

\section{Simulated Images}
In Figure 2, we show simulated images of a gravitational lensing system using the lens parameters of MG0414+0534, where a source image and a lensing galaxy are at redshifts of 2.64 and 0.96, respectively. We assume an SIS profile for the density distribution of a subhalo.
\begin{figure}[!hbt]
\plotone{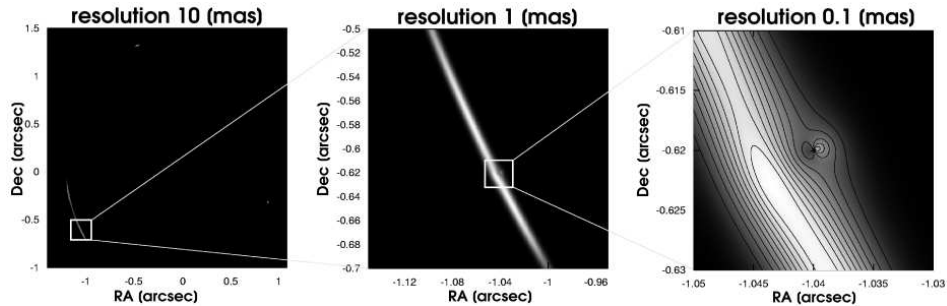}
\caption{Simulated lensed images perturbed by a subhalo with $M_t = 5 \times　10^7 M_\odot$ and $r_t = 1.0$ kpc, where the angular resolution is assumed to be 10, 1 and 0.1 [mas] for left, middle, and right panels, respectively. In this lensing system, a source image and a lensing galaxy are at redshifts 2.64 and 0.96, respectively.}
\end{figure}

The locally distorted feature for an arc-like lensed image in Figure 2 can be caused by not only subhalo lensing but also intrinsic substructure within a source image. To distinguish the effect of subhalo lensing alone, we should select a lens system showing multiple images as a target: intrinsic substructure in a source would be seen in all multiple images, whereas subhalos yield particular features in each image. Thus it is possible to extract the lensing signals of subhalos by subtracting commonly-observed features in lensed images.

\end{document}